\def\year{2022}\relax
\documentclass[letterpaper]{article} 
\usepackage{aaai22}  
\usepackage{times}  
\usepackage{helvet}  
\usepackage{courier}  
\usepackage[hyphens]{url}  
\usepackage{graphicx} 
\usepackage{natbib}  
\usepackage{caption} 
\DeclareCaptionStyle{ruled}{labelfont=normalfont,labelsep=colon,strut=off} 
\usepackage{algorithm}
\usepackage{algorithmic}
\usepackage{devanagari}

\usepackage{newfloat}
\usepackage{listings}
\floatstyle{ruled}
\newfloat{listing}{tb}{lst}{}
\floatname{listing}{Listing}
\usepackage{graphicx}
\usepackage{booktabs}
\usepackage{multirow}

\usepackage{color}
\usepackage{ragged2e}
\usepackage{cuted}
\usepackage{etoolbox}

\title{Characterizing Political Campaigning with Lexical Mutants on Indian Social Media}

\author{Shruti Phadke, \textsuperscript{\rm 1}, Tanushree Mitra, \textsuperscript{\rm 2}}
\affiliations{\textsuperscript{\rm 1} University of Texas at Austin, \textsuperscript{\rm 2} University of Washington}
\usepackage{bibentry}

\begin{document}

\maketitle

\begin{abstract}
Increasingly online platforms are becoming popular arenas of political amplification in India. With known instances of pre-organized coordinated operations, researchers are questioning the legitimacy of political expression and its consequences on the democratic processes in India. In this paper, we study an evolved form of political amplification by first identifying and then characterizing political campaigns with lexical mutations. By lexical mutation, we mean content that is reframed, paraphrased, or altered while preserving the same underlying message. Using multilingual embeddings and network analysis, we detect over 3.8K political campaigns with text mutations spanning multiple languages and social media platforms in India. By further assessing the political leanings of accounts repeatedly involved in such amplification campaigns, we contribute a broader understanding of how political amplification is used across various political parties in India. Moreover, our temporal analysis of the largest amplification campaigns suggests that political campaigning can evolve as temporally ordered arguments and counter-arguments between groups with competing political interests. Overall, our work contributes insights into how lexical mutations can be leveraged to bypass the platform manipulation policies and how such competing campaigning can provide an exaggerated sense of political divide on Indian social media. 

\end{abstract}

\noindent 

\section{Citation}
To cite: Shruti Phadke and Tanushree Mitra. 2024. Characterizing Political Campaigning with Lexical Mutants on Indian Social Media. AAAI International Conference on Web and Social Media (ICWSM 2024), (accepted May 2023).

\section{Introduction}
Online political activism is rapidly becoming a weapon of mass influence on Indian social media. Starting from the Indian general elections in 2014, social media has been used as a campaigning arena in significant political events over the last several years \cite{sohal2018content,akbar2021misinformation,dash2021divided}. 
Especially given the recent allegations \cite{Indiaobj96online,Manipula39online,rajgarhia2020media} of platform manipulation by political parties in India, it is important to invest in computational research that studies online influence operations outside of the West. In fact, a recent study uncovered organized political influence where party supporters received tweet templates through WhatsApp and Google docs and were encouraged to create copypasta campaigns to influence public opinion \cite{jakesch2021trend}. One message quoted in their study also suggests that users may be instructed to tweak the messaging template without directly copy-pasting on social media [\cite{jakesch2021trend} Pg. 9]. 

\begin{figure}[h]
    \centering
    \includegraphics[width=0.95\linewidth]{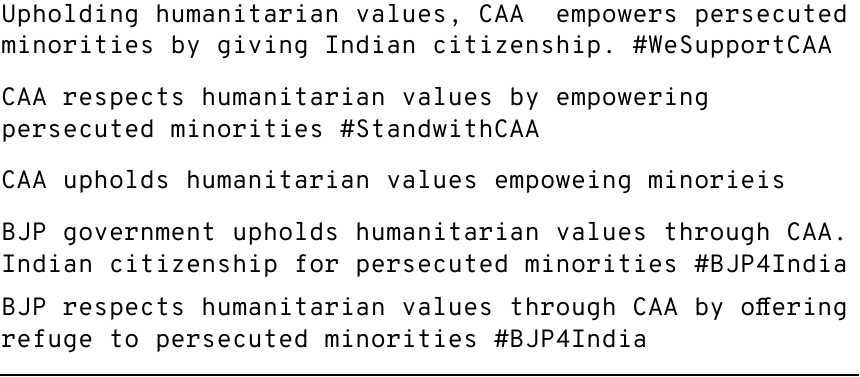}
    \caption{Examples of messages from an amplification campaign with lexical mutants. There are 932 more messages similar to this in different languages (English, Hindi, Punjabi, and Marathi),  distributed by different Twitter accounts and Facebook groups.}
    \label{fig:hiddenampExample}
\end{figure}

For example, consider messages in Figure \ref{fig:hiddenampExample} about a controversial amendment to Citizenship Act in India that offered asylum to religious minorities to surrounding countries, excluding the Muslim population. All messages have the same underlying content, however, each message varies lexically. We argue that subtle lexical mutations such as this, especially spanning over multiple languages, could contribute to a larger political amplification campaign which is harder to detect than a simple copypasta \footnote{a pattern of different users posting identical messages to possibly amplify a certain opinion \cite{weber2021amplifying}}. This paper focuses on detecting and analyzing political amplification campaigns with lexical mutations. By ``amplification campaigns'', we mean coordinated efforts at amplifying certain opinions on online social media \cite{weber2021amplifying}. While analyzing amplification, we specifically study coordination in language through lexical mutants. We use the term \textit{``lexical mutant''} (re-purposed from Meme mutation \cite{simmons2011memes,leskovec2009meme}) to mean content that is reframed, paraphrased, rephrased, or altered while preserving the same underlying message. Moreover, we extend this definition to incorporate mutations across Indian languages. To first find evidence of and then assess the extent of amplification campaigns with lexical mutations, we ask:

\noindent \textbf{RQ1: }How can we identify political message amplification campaigns with lexical mutants on Indian social media?

We focus on two recent national events in India: the introduction of the Farm Bills and the Citizenship Amendment Act CAA (explained in detail in the Data section) and curate a cross-platform dataset of Tweets and Facebook group posts. Considering the use of multiple languages along with English on Indian social media, we use multilingual sentence embeddings with subsequent network analysis and identify over 3.8K political amplification campaigns with lexical mutations. By further establishing that nearly 34\% of the messages in amplification campaigns are unique lexical mutants, our work provides an essential context into the actual expanse of political amplification beyond copypasta.

After identifying the amplification campaigns, we further characterize the use of political amplification across various dimensions. Given that most of the previous studies focus only on one right-wing political party BJP or, primarily focus on only one social media platform, Twitter, we analyze amplification across multiple Indian political parties and extend our analysis to Facebook. 

\noindent \textbf{RQ2: }What are the characteristics of amplification campaigns? 

\hspace{2pt} \noindent \textbf{RQ2a: }How are political messages amplified across the Indian political spectrum?

\hspace{2pt} \noindent \textbf{RQ2b: }How are political messages amplified across different platforms?

\hspace{2pt} \noindent \textbf{RQ2c: }What dominant claims are amplified by accounts with various political leanings?

By manually assigning political leaning to accounts, we find that BJP, anti-BJP, and the accounts supporting other political parties (INC and AAP) all use political amplification equitably (Figure \ref{fig:rq2results}). Furthermore, we find that 30\% of the amplification campaigns spread across platforms (Figure \ref{fig:rq2results} (c)). More interestingly, by analyzing prominently amplified political claims, we find that most large-scale campaigns are reactions to claims made by the political opposition. For example, political claims amplified by anti-BJP accounts are later countered through a similar amplification campaign by BJP (Figure \ref{fig:reactionary}). 


Overall, our results indicate that studies on political influence in India can benefit from a larger political context than focusing on only one party. Moreover, our findings indicate that amplification beyond a simple copypasta may thrive in exerting inauthentic influence on the mass in the absence of updated platform manipulation policies. Finally, our results also contribute a novel understanding of the reactionary landscape of online Indian political influence which can compound radicalization and reduce trust in genuine political expression online.

\section{Background and Related Work}

\subsection{Online Political Influence in India}
After the 2014 general elections, social media has emerged as an important battleground in Indian politics \cite{ahmed20162014,jakesch2021trend,jaffrelot2015modi}. In an attempt to reach out to the younger population, political parties have started deploying organized political campaigns on social media \cite{Election87Indiaonline}. Researchers have studied social media manipulation in India during elections \cite{ahmed20162014,das2021online,jakesch2021trend,jaffrelot2015modi,sohal2018content}, and various civil conflicts such as farmers' protests,
COVID-19 crisis \cite{akbar2021misinformation,dash2021divided} and protests against Citizenship Amendment Act (CAA) \cite{edingo2021social}.

These studies independently observe that  during elections or civil unrest events, social media in India was flooded with various influential narratives promoting political propaganda. For example, 
during the early months of COVID-19, the issues related to the pandemic were used to frame anti-Muslim disinformation and populist narratives \cite{akbar2021misinformation}. 
These disinformation campaigns reflect how political influence narratives in India are closely related to religious fundamentalist attitudes. In fact, an in-depth interview study of supporters of different political parties in India, including the BJP, INC, and Communist party, found that Indian social media users are concerned with an increasing amount of religious fundamentalist appeals and narratives on social media \cite{das2021online}.

\begin{figure*}[t]
    \centering
    \includegraphics[width=0.95\textwidth]{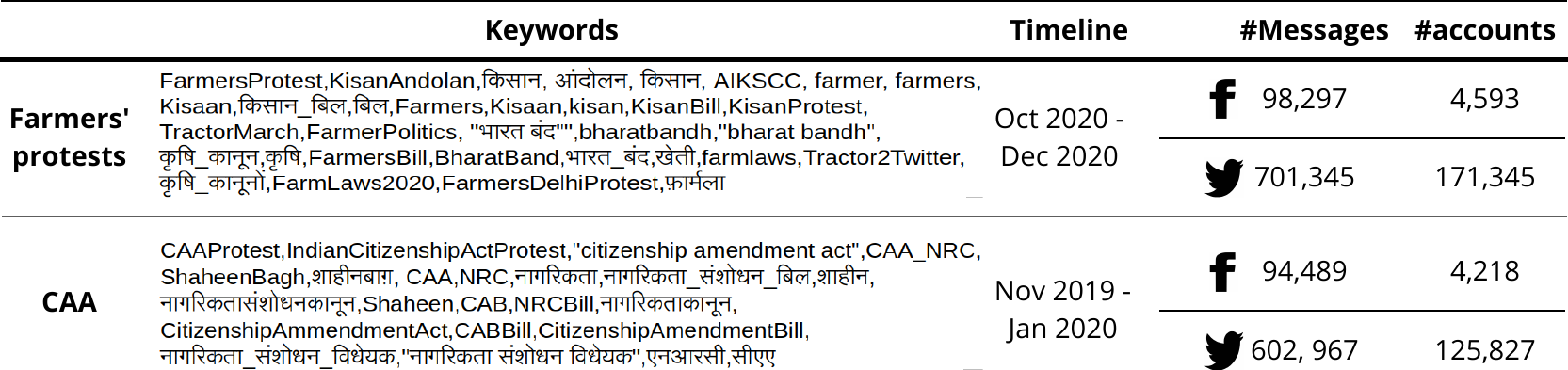}
    \caption{Table describing dataset  keywords, timeline, and the number of posts. Note that we collect only original tweets and posts, excluding retweets and reshares.}
    \label{fig:eventDataInfo}
\end{figure*}

\subsection{Computational Research on online political amplification}
While the presence of influential political rhetoric is evident in Indian social media, computational research exploring coordinated influence operations in India is limited. Jakesch et. al. study 75 copypasta hashtag manipulation campaigns across 600 WhatsApp groups and political Twitter accounts in India and find evidence of centrally controlled political influence operations that benefit from the voluntary participation by the party-followers \cite{jakesch2021trend}. 
While their study brings to light an important phenomenon of influence operations in India, it also motivates further research into identifying such coordinated message templates at scale across more popular platforms such as Facebook. Moreover, while previous research has primarily focused on the ruling political party BJP, it is also important to understand the prevalence of such coordinated influence operations across multiple political parties in India. 
Furthermore, despite Facebook being the most popular social media platform in India \footnote{\url{https://www.statista.com/statistics/1115648/india-leading-social-media-sites-by-page-traffic/}}, most prior works on influence operations in India have focused on Twitter \cite{ahmed20162014,dash2021divided,edingo2021social,jakesch2021trend}. Given the increasing engagement seen by polarizing Indian political Facebook groups \cite{InIndiaF92online}, it is important to investigate the coordinated influence activity on this platform. 

Outside of India, researchers have studied political amplification in the West, by focusing on re-tweet or re-share or co-tweet networks \cite{gallagher2021sustained} or by considering other posting characteristics, such as posting time, user similarity, user coordination in an ensemble \cite{cinelli2022coordinated,hristakieva2022spread,weber2021amplifying}. Most similar to our work is research by Pacheco that investigates White Helmet coordinated influence networks using text similarity based on pattern recognition \cite{pacheco2020unveiling}


In this work, we extend the prior works by first, analyzing political amplification across multiple political parties and languages in India and second, by including previously understudied Indian Facebook. We further include lexical mutants in amplification campaigns that go beyond simple copypasta and consider organic content tweaking that could be used to evade detection from moderation against platform manipulation. In the next section, we start by describing our cross-platform dataset centered around political events in India. 



\section{Data}
\subsection{\textbf{Political events and keywords}} 
In this paper, we study the political message amplification campaigns on Indian social media. To contextualize our analysis, we focus on two recent politically divisive events that affected the whole nation and attracted partisan debate on social media. We used keywords and hashtags related to these events to collect data from Facebook and Twitter. 

\subsubsection{\textit{Farmers' protests: }} Farm acts, passed in the parliament of India in September 2020, sparked nationwide protests by farmers' organizations. Farm unions were primarily protesting the entry of corporations in crop trading facilitated by the farm acts along with legacy issues such as high farmer suicide rates and low agricultural income in India. The protests were highly politicized across the political spectrum in India with the ruling political right (BJP) standing in support of the farmers' bills while the oppositional left (INC, AAP, etc.) aligned with farmers' unions in the protests. 

\subsubsection{\textit{Citizenship Amendment Act (CAA): }} This amendment to the Citizenship Act was proposed by the Government of India under the leadership of the right-wing BJP party. It offered a pathway to Indian citizenship for persecuted religious minorities from Afghanistan, Bangladesh, and Pakistan who are not Muslims. The amendment was opposed by non-BJP politicians and student organizations, causing polarizing tensions across the nation which led to violent protests and rallies from both sides.  

\subsubsection{Selecting keywords for data collection: }
We start building our list of keywords from previous works \cite{dash2021divided,dash2022narrative} around the Farmers' protests and CAAs. We aim to make the keyword list more generic to increase the coverage of the dataset, while still ensuring that we capture the relevant data. For example, we strategically include general words like ``Kisan'' or ``{\dn Esee}'' which are transliterations of the words ``farmers'' and ``CAA'' and are more likely to be used only in the Indian context. Table \ref{fig:eventDataInfo} records the keywords used in collecting the data. 

\subsection{\textbf{Facebook dataset}} 
Facebook groups have been known to play key roles in recruiting supporters for politicians and political parties during elections \cite{woolley20102008}. In fact, Facebook cyber security expert commented that during the 2019 Parliamentary elections in India, Facebook groups and pages were designed to look independent but were actually linked to political parties trying to conceal their identities \cite{HowaReli35online}. Such Facebook groups were also found to be linked with fake or bot accounts that spread misinformation and influential content relevant to elections \cite{InIndiaF92online}. 

We use CrowdTangle's post search API\footnote{\url{https://github.com/CrowdTangle/API/wiki/Search}} to extract posts made on Facebook pages and groups relevant to the political events described above. We use the list of keywords and hashtags mentioned in Figure \ref{fig:eventDataInfo} to collect public posts from Facebook groups and pages. Note that CrowdTangle or any official Facebook API does not offer any authorship information for the posts. 

\subsection{\textbf{Twitter dataset}} We also collect tweets relevant to the two political events using the Twitter Academic Research API. Similar to the Facebook dataset, to analyze hidden amplification campaigns, we only preserve the original tweets, excluding retweets and quote tweets. Overall details of the dataset are available in Figure \ref{fig:eventDataInfo}. 

\begin{figure*}[]
    \centering
    \includegraphics[width=0.99\textwidth]{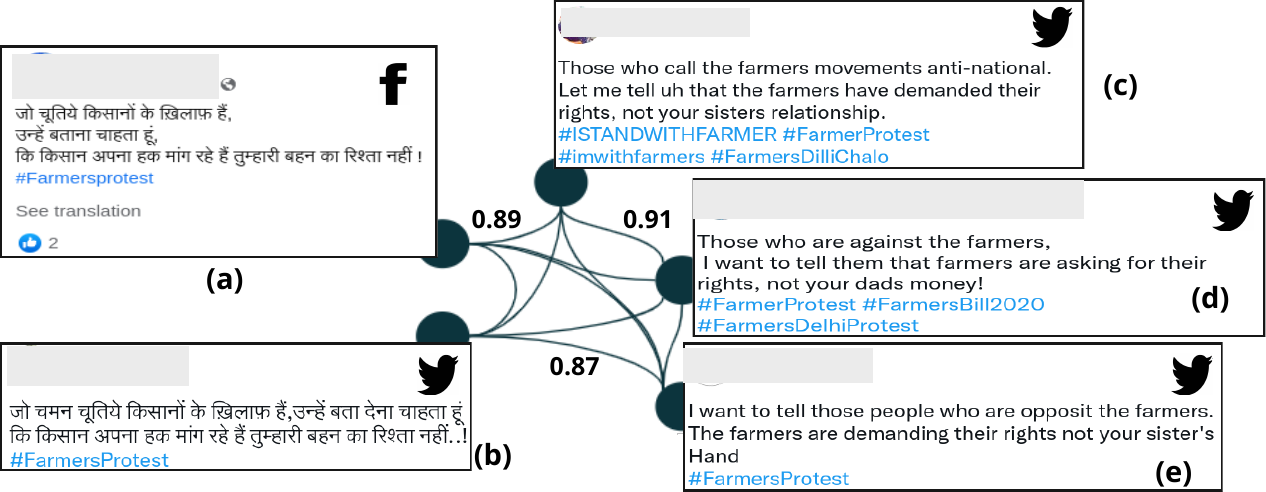}
    \caption{Examples of lexical mutants in an amplification campaign. Two messages (nodes) are connected together if they have high cosine similarity. A clique of nodes connected like this represents an amplification campaign with lexical mutants. }
    \label{fig:psotgroupexample1}
\end{figure*}

\section{RQ1: Identifying Political Campaigns with Lexical Mutations}
A key to identifying amplification campaigns beyond copypasta is to identify messages that are similar to each other in terms of core content but are not explicit re-posts or re-tweets \cite{pacheco2020unveiling}. Unlike retweets or reposts, messages with similar content are treated as different by the platform and the link between the original and the copy is hard to detect \cite{pacheco2020unveiling}. Consider, for example, the messages in Figure \ref{fig:psotgroupexample1}. All of the messages have similar content with slight lexical variations, posted by different users on different social media platforms. Below, we outline our methodology for identifying groups of messages with lexical mutations across multiple languages on Indian social media. 

\subsection{Characterizing post similarity across languages}

\subsubsection{\textit{Extracting multilingual embeddings: }}
A usual approach to identifying messages with slight lexical mutations may be to compare the edit distanced \cite{leskovec2009meme}, message keywords \cite{schoch2022coordination} or to use sentence embeddings with cosine similarity \cite{pacheco2020unveiling}. However, posts on Indian social media contain multiple languages. For example, in Figure \ref{fig:psotgroupexample1}, all the messages have similar content but some are in different languages. 
In this case, simple token-based word embeddings trained in English, or other fuzzy matching methods will not work. 

Instead, we use multilingual sentence embeddings to represent texts across multiple Indian languages and also capture the semantic similarity between paraphrased or reframed texts. There are several pre-trained multi-lingual models available, such as sentence-BERT \cite{reimersdistill2020}, Language-Agnostic SEntence Representations (LASER) \cite{artetxe2019massively} and multilingual BERT \cite{devlin2018bert}. LASER was found to outperform multilingual BERT for Hindi text classification \cite{joshi2020deep}. Since a significant portion of our text data is in Hindi, we use the LASER model which works with more than 90 languages containing more than 28 different kinds of alphabets. LASER is optimized and evaluated for parallel sentence extraction and translations across different languages, making it suitable for characterizing cross-lingual similarities \cite{artetxe2019massively}. 

\subsubsection{\textit{Identifying similarity threshold for lexical mutation: }} Similarity between two texts can be calculated with the cosine similarity between the multilingual sentence embeddings described above. A cosine similarity of 1 indicates perfect similarity between the two texts. Examples in Figure \ref{fig:psotgroupexample1} all contain a similar underlying message with variations in languages and phrases. To extract groups of texts such as this, we need to determine a threshold value of cosine similarity above which we can consider two texts to be lexical mutants of each other. We determine this threshold empirically by manually analyzing pairs of texts with different cosine similarity scores. Specifically, we randomly sampled 20 pairs of sentences for cosine similarity values each, starting from 0.5, with increments of 0.05. Some of the sampled pairs contained texts from the same language whereas other pairs contained texts from different languages. We labeled each pair as either 1---to indicate whether a lexical mutation still resulted in preserving the underlying message---or 0. The number of pairs scored as 1 in the sample of 20 naturally kept increasing with the cosine similarity score. All samples with a score of 0.85 were labeled as 1. Hence, we chose 0.85 as the cosine similarity threshold. In other words, for the rest of the downstream analysis, ``lexical mutant'' posts will mean posts with cosine similarity between multilingual embeddings equal to or above 0.85. 

\subsection{Finding amplification campaigns with lexical mutations}
After identifying lexical mutants through pairwise similarity of multilingual embeddings, to analyze amplification campaigns on a larger scale, it is important to determine large groups of similar messages. For example, there are several other messages similar to the examples in Figure \ref{fig:psotgroupexample1} with high semantic similarity to each other. How can we find large groups of lexical mutants? We next take a network-based approach to identify groups of lexical mutants. 

\begin{figure}[t]
    \centering
    \includegraphics[width=0.95\linewidth]{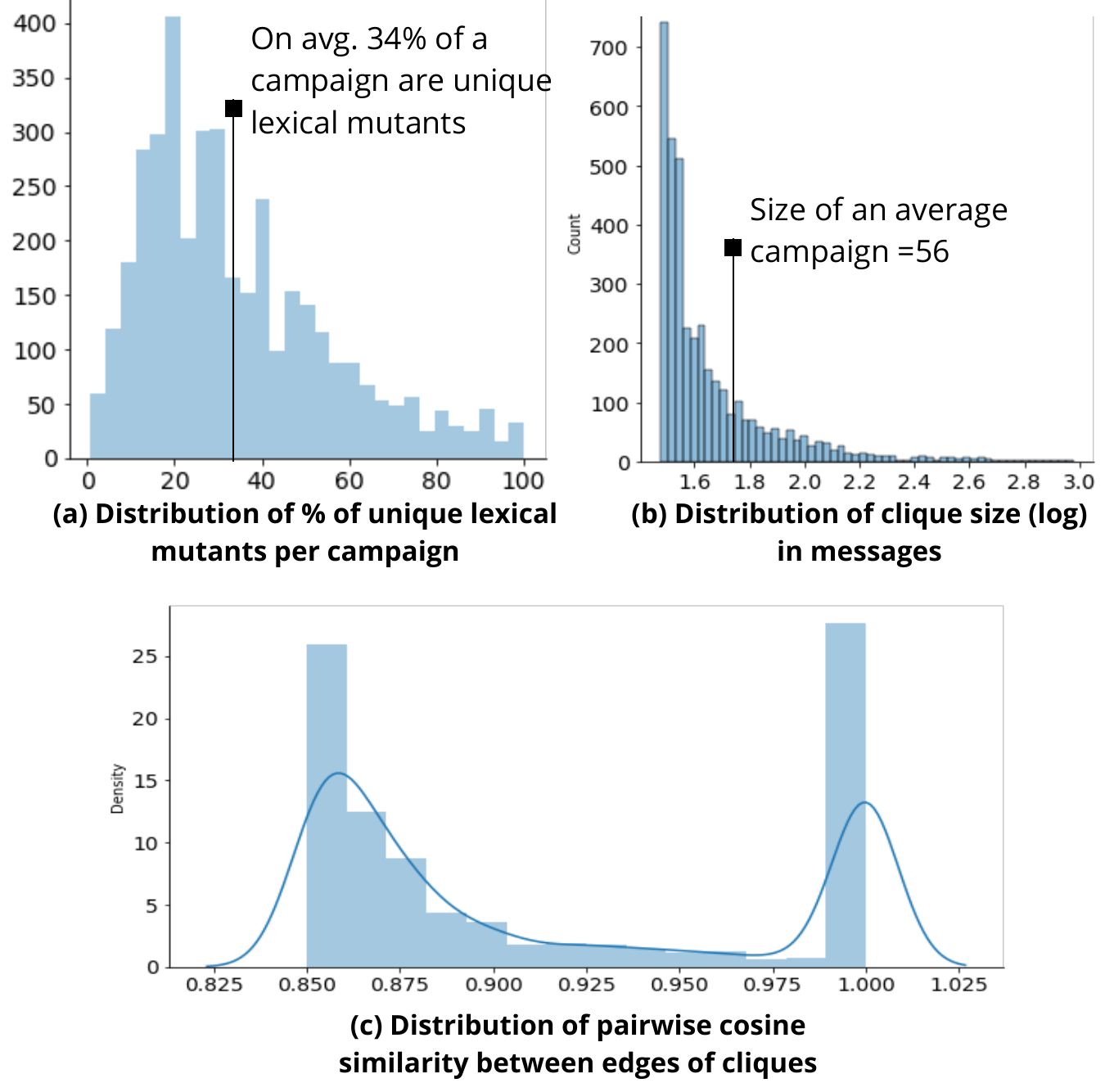}
    \caption{(a) Represents the distribution of  the percentage of unique lexical variants in each campaign, calculated by removing all duplicated texts. (b) Shows the distribution of clique sizes in log scale. The average size of the clique is 56 messages. (c) Displays the KDE plot for pairwise cosine similarities between pairs of texts included in cliques. The two peaks of bimodal distribution represent our empirically determined cutoff of cosine similarity (left peak) and also the duplicated texts without lexical variations with the perfect cosine similarity (right peak). }
    \label{fig:clique_dig}
\end{figure}

\subsubsection{\textit{Network of texts: }}

Two messages are connected with each other through an edge if they have a cosine similarity equal to or more than 0.85 between their multilingual embeddings. In a network such as this, finding groups of lexical mutants will be equivalent to finding clusters of nodes that are all connected to each other through high cosine similarity. In other words, clusters of completely connected nodes will represent texts which are all lexical mutants of each other. This is commonly referred to as finding cliques \cite{clark1991first}. A clique is a sub-graph in which all nodes are connected to each other through an edge (example Figure \ref{fig:psotgroupexample1}). In a network of messages connected with high cosine similarity, all similar messages will form a clique.

\subsubsection{\textit{Finding lexical mutants through cliques: }}
Computing cliques in large networks is a computationally expensive task. However, the method in which our network is constructed---drawing edges between nodes only when there is high cosine similarity---allows for a large number of connected components. For example, in the Farmers' protest dataset, we had a total of 799K messages (nodes). Out of which 301,342 nodes had at least one edge. The resulting network had around 2.1K connected components. Connected components make for completely disjoint subgraphs and provide a much more computationally affordable space to find cliques. In fact 1,857 of the total components were also perfect cliques, suggesting that connected components could be a good approximation for finding cliques in a network such as this. A similar approach has proven successful in finding coordination networks in White Helmets \cite{pacheco2020unveiling}. We consider each clique found with this method as a single campaign with lexical mutants.

\subsection{Results: Lexical mutant amplification campaigns}
In total, we find 2,558 amplification campaigns relevant to Farmers' protests spanning over 231,896 messages and 1,268 campaigns in the CAA dataset spanning over 146,465 messages. In the results, we only include campaigns with at least 10 messages with no two messages shared by the same Twitter handle, Facebook group, or page. While it is easier to identify and exclude multiple similar messages posted by a Twitter handle, it is challenging to establish this in the Facebook data, given that we dont have user-level information on Facebook groups or page posts. For every clique, we only consider one message per Facebook group or page. It is possible that we are still considering messages posted by the same user in different Facebook groups, inflating the sizes of the cliques. To find out the extent of this, we manually analyze the authorship distribution for Facebook data points in a random sample of 1000 cliques. We found that 78\% of the Facebook messages in a clique come from different users.  

An average amplification campaign has 56 messages posted by different users (Figure \ref{fig:clique_dig} (b)). We found 219 campaigns with at least 100 messages from different user accounts and the largest campaign in the dataset contains 1,232 messages. We find that 29\% of all messages in the Farmers' protests dataset and 21\% of all messages in the CAA dataset were part of an amplification campaign. Moreover, on average, 34\% of the messages in the amplification campaigns are unique lexical mutations (Figure \ref{fig:clique_dig} (a)). The rest of the messages are copypastas of different lexical mutants, indicating that including lexical mutants helps in identifying larger amplification campaigns than simple copypastas. 

\begin{figure*}[th]
    \centering
    \includegraphics[width=0.55\textwidth]{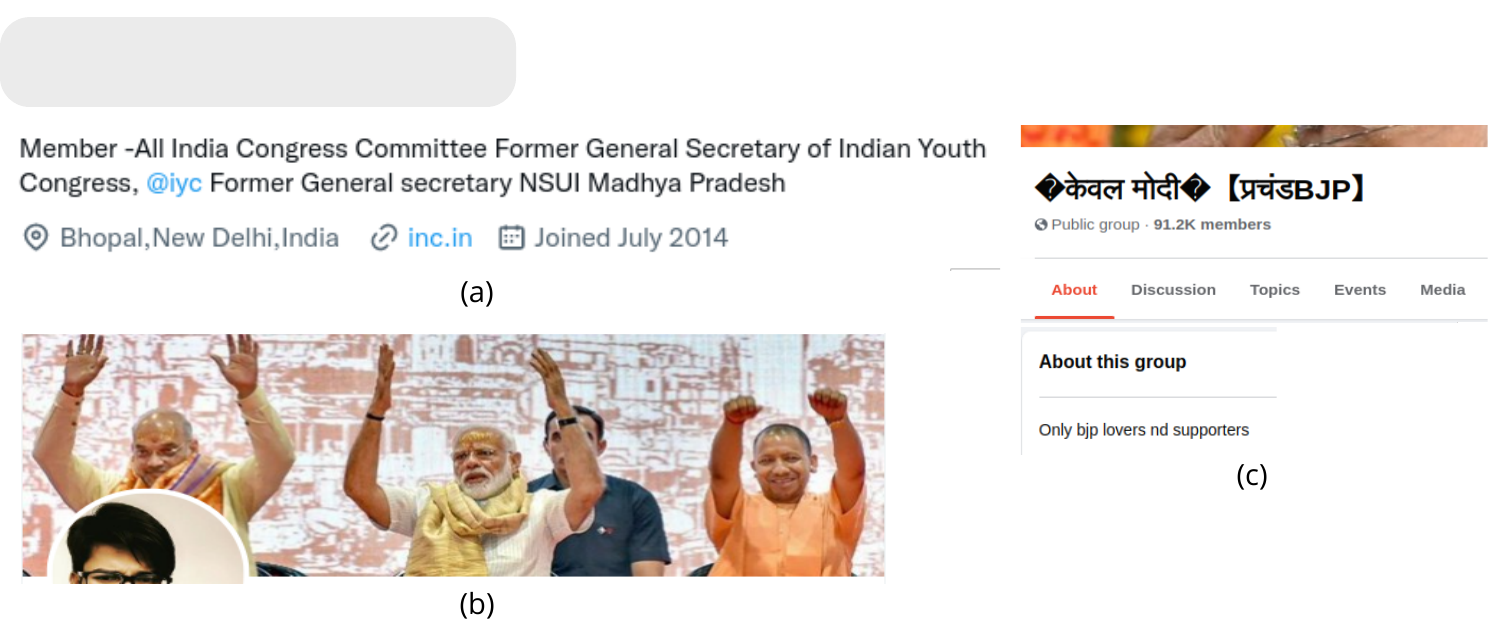}
    \caption{Examples of account bio, pictures, and descriptions in the dataset }
    \label{fig:annotation_Account_exampels}
\end{figure*}
\subsubsection{\textit{Evaluating identified amplification campaigns: }}
How accurately does the methodology described above identify political amplification with lexical mutations? To evaluate, we analyze a random sample of 200 cliques (5\% of all the cliques found) in the dataset and label for the similarity of the messages in the clique. All randomly sampled cliques were labeled by two annotators. We use a conservative labeling scheme--- 1 if \emph{all} messages in the clique satisfy our definition of lexical mutations (see Introduction) and 0 if even one message is semantically dissimilar to the rest of the clique. Particularly, we labeled a clique as 1 only if \emph{all} the messages in the clique contained the same underlying claim. Specifically, we looked for the similarity of topic, opinion, and argumentation while assessing the messages. 

The annotators agreed on 197 out of 200 clique labels, resulting in 98.5\% agreement. For final labels, we  marked cliques as '1' where \emph{both} annotators agreed on the label. We consider all disagreements as '0'. With this adjustment, we find that 191 cliques out of 200 sampled (95.5\%) satisfy our criteria for lexical variants, indicating that our methods can identify amplification campaigns with lexical mutants with high confidence. 

\section{RQ2: Characterizing hidden amplification campaigns}
In RQ1, we identified 2,558 amplification campaigns relevant to Farmers' protests and 1,268 campaigns around Citizenship Amendment Act (CAA). 
While the previous literature has largely focused on studying political influence by the Indian right-wing, in this question we examine the extent of hidden amplification across the political spectrum in India. Moreover, we also analyze the expanse of political amplification across platforms 
 and evaluate dominant narratives. Toward this goal, we first identify the political leaning of the accounts involved in the amplification campaign and measure the use of political amplification on Facebook and Twitter across different parties. We start by outlining our process for identifying the political leanings of social media accounts.

\subsection{Labeling accounts with political leanings}
The amplification campaigns detected in the earlier research questions, spread over nearly 40K social media accounts. Currently, there is only one large-scale dataset---NivaDuck---by Panda \cite{panda2020nivaduck} that records the political leanings of 18,500 Twitter accounts. However, only 368 accounts from NivaDuck overlap with the accounts in our dataset. Hence, we manually analyze the accounts in our dataset and record their political leanings. To better manage labeling resources, we focus only on repeat offenders---accounts that repeatedly participate in different amplification campaigns. In total, we recorded leanings of 493 Facebook groups or pages and 1,631 Twitter accounts that were involved in 5 or more amplification campaigns. We want to note that for most cases, specifically for  97.2\% Facebook accounts and 87.3\% Twitter handles, the accounts had clear self-declarations of their political leaning, leaving little room for interpretation. For the rest of the accounts, we used a combination of manual and quantitative analysis to infer the political leanings. We describe both of these steps in detail next.

\subsubsection{\textit{Labeling accounts based on metadata: }}

\begin{itemize}
    \item \textbf{Twitter: } We first cross-referenced the Twitter handles with the NivaDuck dataset \cite{panda2020nivaduck} and borrowed the labeles readily available. To label the remaining Twitter handles, we first look for cues in the Twitter bio,  username, profile image, and background image for explicit political party affiliation. For example,  the Twitter profile in Figure \ref{fig:annotation_Account_exampels} (a) explicitly declares an association with the Indian National Congress (INC) political party. In some cases, the accounts  also signal political affiliation through profile or Twitter background pictures, as shown in \ref{fig:annotation_Account_exampels} (b). We were able to record definitive leanings of 87.3\% of Twitter handles through the account metadata. 

    \item \textbf{Facebook: } Assigning political leanings to Facebook groups of pages was significantly easier than Twitter. Our dataset contains 493 Facebook groups and pages that host messages used in at least 5 or more amplification campaigns. We label political leanings based on the group's name, description, rules, and recent posts. We find that in most cases, Facebook groups and pages had a clear political affiliation signaled either through group name or description. For example, the Facebook group in Figure \ref{fig:annotation_Account_exampels} (C) is restricted only to the BJP supporters. Over 97.2\% of Facebook, accounts had clear political affiliations represented in the metadata. Next, we explain our steps for assigning political leanings to the accounts with information beyond the account metadata.

\end{itemize}

\subsubsection{\textit{Labeling accounts using posts and further validation: }}
For the accounts that do not signal explicit political affiliation through metadata, we read through the 20 most recent tweets or Facebook posts and also consider the messages in the amplification campaigns associated with the account. Specifically, the first author of this paper looked at whether the accounts posted content supporting specific political parties or politicians, policies, or events relevant to specific political parties. We were able to assign leanings to all of the remaining Facebook groups and 182 of the 207 remaining Twitter handles. The unlabeled Twitter handles are removed from the downstream analysis. 

To further validate the manual labeling, we look at the accounts' network in amplification campaigns. In this network, two accounts are connected together if they ever participate in the same amplification campaign. We next use the label propagation algorithm \cite{cordasco2010community} to infer the leanings of the accounts that did not have explicit affiliation in the metadata (See ``Labeling accounts based on metadata''). We further compare the labels generated with label propagation and manual annotation and find 86\% overlap. This can provide for external validation of the manual annotations efforts. We further discuss the challenges and limitations of manual labeling in the limitations section.

\begin{figure*}[th]
    \centering
    \includegraphics[width=0.98\textwidth]{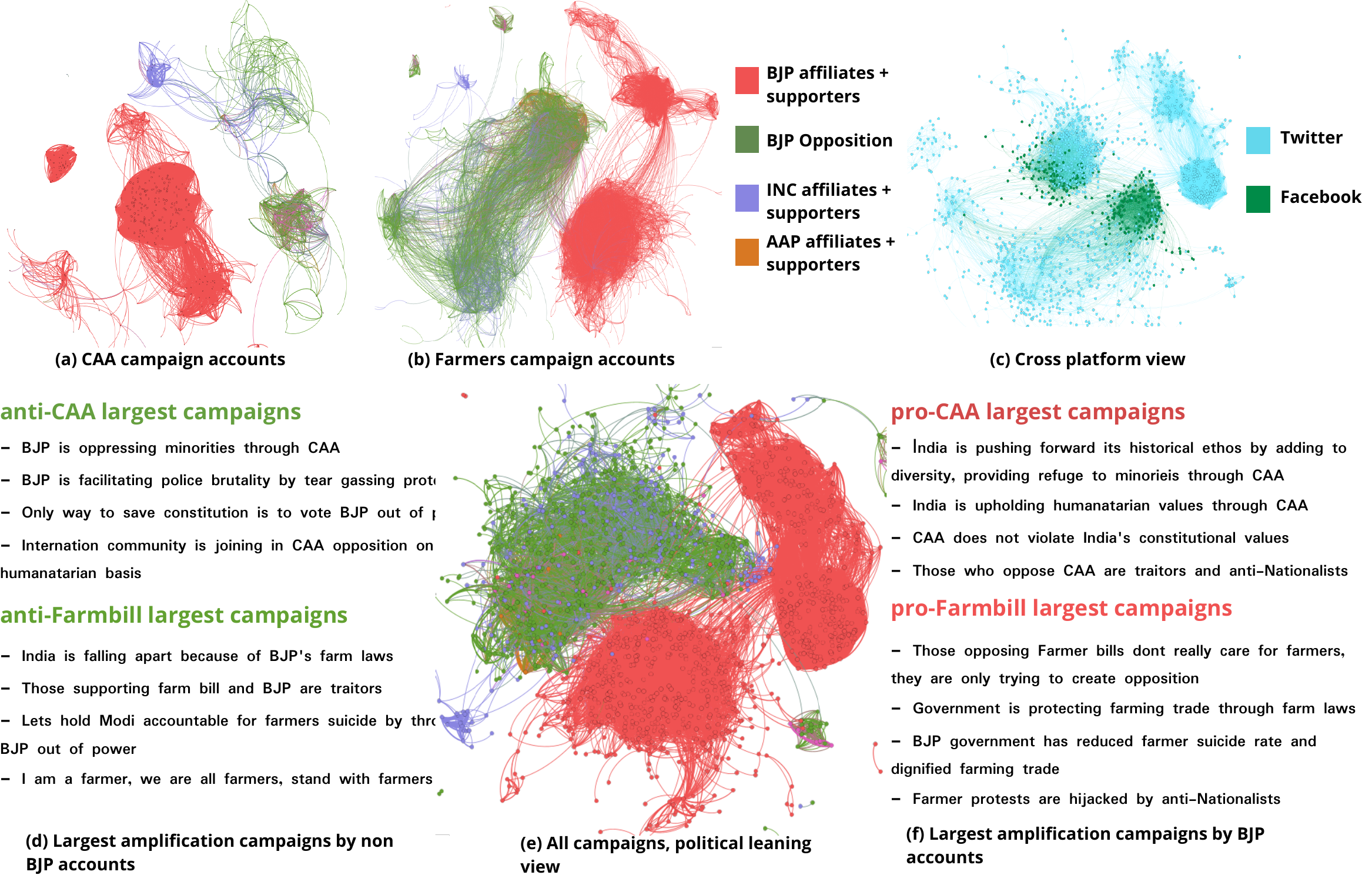}
    \caption{(a) and  (b) represent the network of users involved in political amplification relevant to CAA and Farmers' protests respectively. (e) is the overall user network across both events.  Nodes are the user accounts and colors represent political leaning. The precise percentage of accounts belonging to different political leanings is reported in Table \ref{tab:rq2_stats}. (c) represents the combined network of users across Facebook and Twitter. The node color distinguishes Facebook users (green) from Twitter users (blue). (d) and (f) list messages from the largest amplification campaigns from (d) non BJP accounts and (e) BJP accounts}
    \label{fig:rq2results}
\end{figure*}
\subsubsection{\textit{Political leanings labeling scheme: }} The annotation process described above, resulted in the following labeling scheme. We focus on three political parties---BJP, INC, and AAP---that had the largest number of social media accounts in the NivaDuck dataset \cite{panda2020nivaduck}. We also include \% accounts belonging to each leaning in brackets:

\begin{itemize}
    \item \textbf{BJP affiliates (17\%):} Accounts that explicitly signal affiliation with the ruling right-wing political party. Many of these accounts are created in support of the leading BJP politicians and the 2024 general elections. Widely protested Farmers' bills and Citizenship Amendment Act (CAA) were advocated under BJP leadership and support.  

    \item \textbf{BJP supporters (35\%):} Accounts that did not explicitly affiliate with the BJP but frequently posted content in support of BJP's mission and politicians. 

    \item \textbf{INC affiliates (2\%): } Accounts that openly affiliate with the leading opposition and Indian left-wing party---Indian National Congress (INC). INC joined the general opposition to Farmers' bills and CAA. 

    \item \textbf{INC supporters (10\%): } Accounts that do not clearly affiliate with INC but post content supporting INC politicians and mission. 

    \item \textbf{AAP affiliates (0.6\%): } Accounts affiliated with Aam Admi Party (AAP). AAP is often referred to as the ``third opposition'' with appeals to the common-man identity in India. AAP also joined in the opposition to  Farmers' bills and CAA. 

    \item \textbf{BJP opposition (32\%): } A large chunk of the accounts that while not affiliating with any specific political party, explicitly oppose BJP politicians and policies.

\end{itemize}

\subsection{Results RQ2: Characterisitcs of amplification campaigns}
In the previous analysis, we identified political amplification campaigns around Farmers' protests and CAA and labeled the political leanings of the accounts participating in the campaigns. Here we discuss partisanship, social media platform use, and narratives in the most widespread amplification campaigns. Figure \ref{fig:rq2results} displays various views of the network of users involved in the amplification campaigns. Nodes are user accounts and edges connect two users that are involved in the same amplification campaign. Various network views are colored differently to represent user learning and social media platforms. For example, in (a), (b), and (e) node colors correspond to political leaning while in (c), node colors are based on the social media platform of the account. 

\begin{table*}[]
\centering
\resizebox{\textwidth}{!}{%
\begin{tabular}{@{}ll@{}}
\toprule
\multicolumn{1}{c}{\textbf{\begin{tabular}[c]{@{}c@{}}Core arguments by BJP affiliated accounts\\ (number of large campaigns)\end{tabular}}} & \multicolumn{1}{c}{\textbf{\begin{tabular}[c]{@{}c@{}}Core arguments by non-BJP affiliated accounts\\ (number of large campaigns)\end{tabular}}} \\ \midrule
BJP is fostering diversity and humanitarian values and historical legacy (2) & Modi's policies are anti-humanitarian and shameful for Indians (1) \\
Government is promoting diversity through CAA (1) & BJP is threatening minorities through CAA (1) \\
CAA is designed respecting the Indian constitution (1) & CAA is unconstitutional (1) \\
Government is preserving the culture of northeast India with CAA (1) & Modi is isolating people from northeast India (1) \\
It is the duty of every Indian to support the government we voted for (1) & We should stand up to Modi and his bullying (1) \\
Farmers' protests are disrupting daily businesses and peace (1) & Government is terrorizing peaceful farmers' protests (1) \\
Opposition parties do not care about farm bills. They just want to oppose Modi (1) & Modi supporter or not, anyone who is a patriot should oppose farmers' bills (1) \\
Farmers' protests are hijacked by paid protestors (1) & Farmers dont have IT cells or paid supporters (1) \\
Modi effect: Modi has made farmers' lives better (2) & All kinds of farmers are opposing farm bills because CAA will ruin their lives (1) \\ \bottomrule
\end{tabular}%
}
\caption{Table recording some of the arguments and their counterparts identified through manual analysis of the large (\textgreater 500 messages) amplification campaigns. We identified a total of 9 arguments and counterarguments, spanning 20 distinct amplification campaigns. The left column presents narratives amplified by BJP affiliated accounts and the right column presents the counter-narrative identified through manual analysis. The numbers in parenthesis represent the number of amplification campaigns with the specified core narrative. }
\label{tab:react_arguments}
\end{table*}

\subsubsection{\textit{ RQ2a Political amplification and partisanship: }} Figure \ref{fig:rq2results} (e) represents the network of users involved in political message amplification across both, Farmers' protests and CAA. Overall, 38\% amplification campaigns spread through BJP accounts, 40\% spread through BJP opposition accounts, and 22\% amplification campaigns are propagated through accounts of other parties such as INC and AAP. More granular results across events are present in Table \ref{tab:rq2_stats}. Overall, we find that accounts all across the Indian political spectrum participate in political amplification in an equitable way. 

After analyzing the networks of users that repeatedly participate in the amplification campaigns, we observe that BJP accounts are strongly clustered together without having edges to any other political leanings. On the other hand, non-BJP accounts (BJP opposition accounts with no specific party affiliation, INC accounts, and AAP accounts) all participate in common amplification campaigns.  

\begin{figure*}[th]
    \centering
    \includegraphics[width=0.98\textwidth]{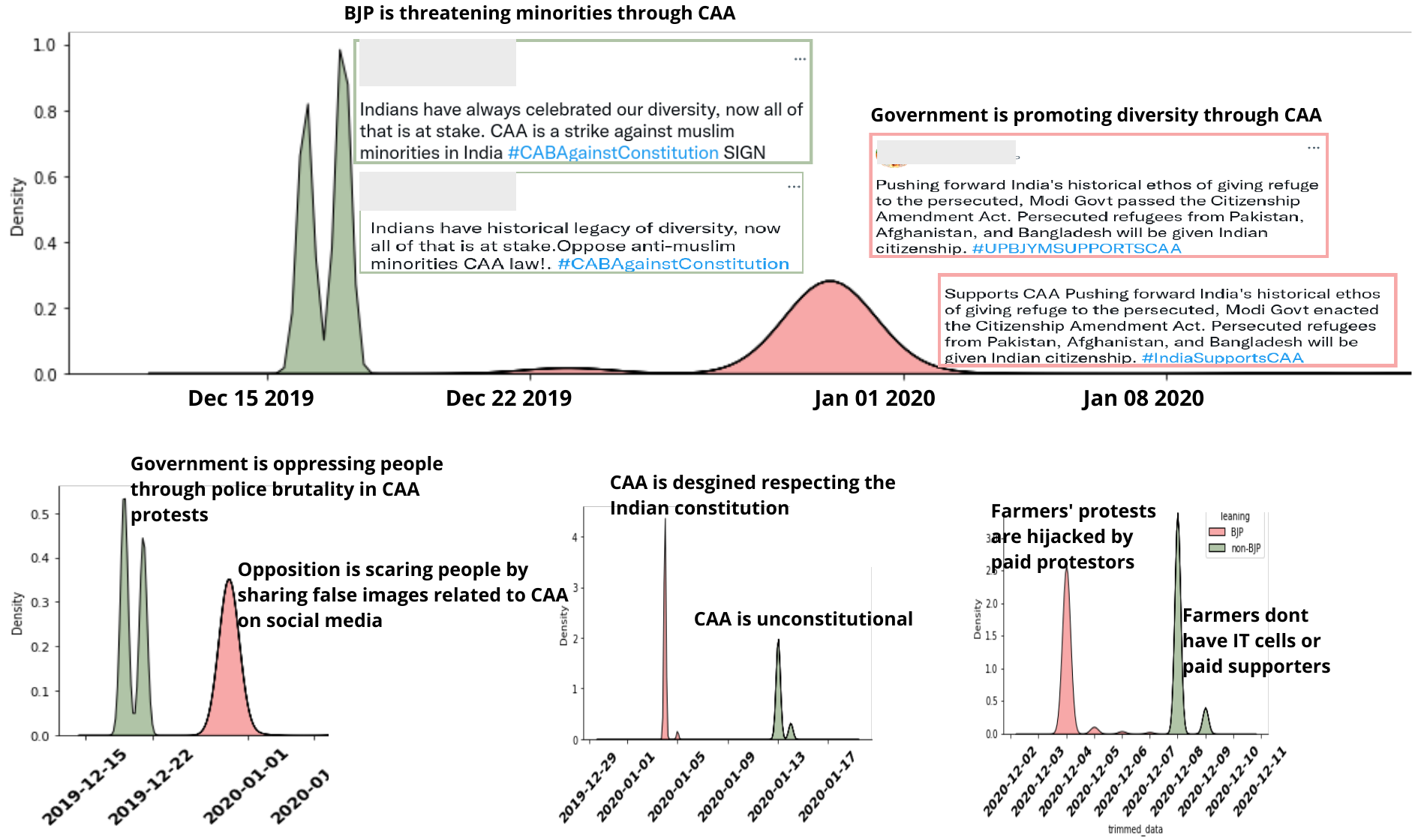}
    \caption{Timelines of largest amplification campaigns. In each plot, x-axis represents the actual calendar dates, and y-axis displays the proportion of messages in the amplification campaign posted on that date. We specifically analyze the timelines of the largest campaigns on the opposite sides of the political spectrum that make contrasting claims.}
    \label{fig:reactionary}
\end{figure*}

\subsubsection{\textit{RQ2b Political amplification across platforms: }} Figure \ref{fig:rq2results} (c) displays the network of user accounts involved in amplification with node color representing the social media platform (blue: Twitter, green: Facebook). In sum, 65\% of the detected campaigns reside only inside Twitter, while 5\% are confined to Facebook. 30\% of the amplification campaigns spread across both platforms.

\subsubsection{\textit{RQ2c Prominent political claims in amplification campaigns: }} We further analyze the dominant claims in amplification campaigns across different political leanings. Some examples of amplification narratives are described in Figure \ref{fig:rq2results} (d) and Figure \ref{fig:rq2results} (e). We manually analyzed the narratives in 63 large amplification campaigns---campaigns with more than 500 messages. Specifically, the first author of this paper manually noted the core message in each of the 63 amplification campaigns by analyzing the contents of the lexical mutants inside the campaign. During this process, we noticed that the largest amplification campaigns by BJP and non-BJP accounts could be considered direct antitheses of each other. For example, the largest anti-CAA campaign by non-BJP accounts claims that BJP is oppressing minorities through CAA. Whereas, the largest amplification campaign by BJP accounts claims that India is continuing their historical legacy of providing refuge to minorities through CAA. Similarly, non-BJP accounts attempt to depict Farm Bill supporters as traitors whereas BJP accounts criticize the opposition to Farm Bill for its political opportunism. 

To further systematically analyze these contrasting political campaigns, we manually paired each amplification campaign with its counterargument, wherever possible. Out of the 63 large campaigns, we found counterarguments for 20 using the core campaign message recorded before. Table \ref{tab:react_arguments} records the core messages of contrasting political amplification campaigns. 


We further analyze the timelines of contrasting campaigns. We find that the claims and counter-claims follow precise temporal orders. For example, the bulk of messages suggesting that BJP is threatening minorities through the CAA (Figure \ref{fig:reactionary}) was posted by non-BJP accounts around December 15th, 2019. Amplification campaigns by BJP contesting this claim were spread during the subsequent days claiming that CAA promotes diversity and the historical legacy of providing refuge to minorities. Similar trends followed in the 17 other contrasting campaigns. Overall, the temporal order political messaging suggests that amplification campaigns are possibly being used as devices to counter the narratives from both sides of the Indian political spectrum.

\begin{table}[h]
\resizebox{0.9\columnwidth}{!}{%
\begin{tabular}{@{}clllll@{}}
\toprule
\textbf{\begin{tabular}[c]{@{}c@{}}accounts \\ from\end{tabular}} & \multicolumn{1}{c}{\textbf{CAA}} & \multicolumn{1}{c}{\textbf{\begin{tabular}[c]{@{}c@{}}Farmers' \\ protests\end{tabular}}} & \multicolumn{1}{c}{\textbf{Overall}} & \multicolumn{1}{c}{\textbf{TW}} & \multicolumn{1}{c}{\textbf{FB}} \\ \midrule
BJP & 47\% & 35\% & 38\% & 37\% & 39\% \\ \cmidrule(l){2-6} 
\begin{tabular}[c]{@{}c@{}}BJP \\ Opposition\end{tabular} & 35\% & 44\% & 40\% & 41\% & 40\% \\ \cmidrule(l){2-6} 
\begin{tabular}[c]{@{}c@{}}Other \\ parties\end{tabular} & 18\% & 21\% & 22\% & 22\% & 21\% \\ \bottomrule
\end{tabular}%
}
\caption{Participation in political amplification by accounts of different political leanings. Here we report on the proportion of accounts with different leanings in different political events and platforms.}
\label{tab:rq2_stats}
\end{table}




\section{Discussion}
In this paper, we analyzed how political parties leverage amplification campaigns with lexical mutations to popularize contrasting political stances on Indian social media. Focusing on two key recent events ---Farmers' protests and the introduction of the Citizenship Amendment Act (CAA)---we first identified over 3.8K amplification campaigns across Facebook and Twitter. Next, we characterize the use of amplification campaigns across multiple political parties in India and multiple and social media platforms.  Our results provide an updated understanding of political amplification by looking beyond the popularly studied platform Twitter and by considering the multi-party political landscape of India. 
Below we discuss some of our findings in detail. 

\subsection{Amplification with lexical mutants: Moving beyond copy-pasta}
Previous studies have revealed the presence of copy-pasta political campaigns both in India \cite{jakesch2021trend} and the West \cite{franccois2019ira}. We focus on surfacing large-scale campaigns that not only include copy-pastes but also detect slight changes in phrasing and language---lexical mutations. The analysis of large amplification campaigns detected in RQ1, clearly indicates that the messages have a similar underlying template. Here we discuss two key questions raised by our findings: do users deliberately tweak the messages for amplification and why? 

\subsubsection{\textit{Hidden amplification and platform manipulation policies: }}In one of the previous studies \cite{jakesch2021trend}, researchers found that local political leaders in India seek out participation through WhatsApp groups to spread messages from ``Tweet banks''. One such message quoted in their paper explicitly instructs users to alter the wording in the template: \textit{``Note-Please don’t just copy-paste the sample tweets, please alter it a bit.''}[\cite{jakesch2021trend} Pg. 9]. Our results indicate that users might be getting these types of instructions on a larger scale for campaigns of different political parties. Moreover, we argue that this kind of hidden amplification might be designed to bypass the platform manipulation and spam policies \footnote{\url{https://help.twitter.com/en/rules-and-policies/copypasta-duplicate-content}}. For example, on Twitter, copypasta campaigns are subject to review, and the platform is committed to reducing the visibility of such content. In fact, repeatedly violating copypasta rules is considered a severe violation of the platform manipulation policy. In our study, we identify thousands of user accounts that repeatedly participate in amplification campaigns (Figure \ref{fig:rq2results}). 
The fact that we were able to detect large-scale campaigns and repeat offender users, who are still active on the platforms, may suggest that hidden amplification may be proving effective in evading platform governance policies. In the future, a more standardized and scalable approach such as the one suggested in RQ1 could be required to counter large-scale platform manipulation. For example, our entire pipeline to detect amplification campaigns with lexical mutants could be used in real-time to detect political amplification. Detecting messages involved in amplification could help platforms limit their visibility soon after they are published, and also keep track of repeat offenders across different political events. 


\subsection{Indian political amplification campaigns and reactionary politics}
Our study offers a unique look into the political influence exerted by accounts from various political leanings in India. By considering the multi-party political spectrum in India, we provide an essential context to previous research focusing on only one political party. More importantly, our results reveal the reactionary nature of political influence in India (Figure \ref{fig:reactionary}). Specifically, by investigating the temporal order of contrasting amplification campaigns, we observe that large-scale amplification campaigns arise in the form of arguments and counter-argument between different political viewpoints. Researchers studying the role of copypasta in reactionary politics around neo-Nazism in the United States propose that such reactionary politics often gives rise to radicalization \cite{topinka2022politics}. Specifically, influencing public arguments and counter-counterarguments through amplification can harm democratic equality and can create an exaggerated sense of democratic divide in the public eye. Our methods can also be extended to study the temporal patterns in political amplification campaigns outside of India. 

\subsection{Ethical Considerations}
Adhering to AAAI code of conduct and ethics guidelines, here we discuss the stakeholders, harm, privacy, and confidentiality in our research. First, we acknowledge that social media users and social media platforms are primary stakeholders in this research. With this work, we intend to contribute insight into both, the expanse of political amplification on social media as well as computational ways to detect political amplification which could be useful in platform governance. This research is retrospective and involves no interaction with the users studied. Moreover, we do not include any user information in the paper and blur out user handles from shared figures. This limits the harm and maintains user privacy. Moreover, all the data used in this paper is publicly available preserving the data confidentiality on social media. 

\section{Limitations}
Our work primarily looks at political amplification campaigns with lexical mutants, without including retweets or quote tweets. Incorporating retweets and other signals, along with lexical variations, or image or video coordination can provide a more comprehensive picture of political amplification in India. Moreover, as mentioned while discussing RQ1 results, we can not establish the authorship of each and every post made in Facebook groups or pages. While only 5\% of amplification campaigns strictly reside inside Facebook, and while our manual robustness check reveals that 78\% of messages in the campaign come from different users, we acknowledge that having authorship information could enhance our results. Currently, there is no way to obtain this information through any official API, and web crawling for author information on Facebook posts could pose serious ethical challenges. While calculating multi-lingual embeddings we used the LASER model, which is trained and evaluated on translated pairs of texts. However, our current methodology does not account for transliterations, which can be incorporated in the future to find larger campaigns. 
Moreover, our process for identifying the narratives and counter-narratives involved only one annotator with proficiency in multiple Indian languages and familiarity with the Indian political landscape. The annotation performance could be made better by involving more expert annotators. Finally, while it is easier to detect which messages are part of a political amplification campaign, we still cannot determine whether the message reflects the user's genuine political opinion. 

\section{Conclusion}
In this work, we analyze Indian political amplification campaigns with lexical mutations. By modeling messages in national political events, we uncover over 3.8K political amplification campaigns across languages, social media platforms, and political parties in India. Our findings enable future research that can explore the repeat offender accounts in-depth and assess their legitimacy.

\bibliography{aaai22}


\end{document}